\documentclass[prd,preprintnumbers,nofootinbib,aps]{revtex4-1}
\usepackage{graphicx}
\usepackage{dcolumn}
\usepackage{bm}
\usepackage{epsfig,amsmath}
\usepackage[usenames,dvipsnames]{color}
\usepackage[pagebackref=false, colorlinks=true]{hyperref}
\usepackage{appendix}
\usepackage{subfigure}
\definecolor{redish}{rgb}{0.7,0.2,0.0}  
\definecolor{bluish}{rgb}{0.2,0.5,0.8}

\hypersetup{linkcolor=redish,          
                  citecolor=blue,        
                  filecolor=magenta,      
                  urlcolor=bluish}          

\DeclareFontFamily{U}{rsfs}{}         
\DeclareFontShape{U}{rsfs}{m}{n}{<5> rsfs5 <6><7> rsfs7          %
  <8><9><10><10.95><12><14.4><17.28><20.74><24.88> rsfs10}{}     %
\DeclareMathAlphabet{\mathfs}{U}{rsfs}{m}{n}                     %

\newcommand{\ba}{\nopagebreak[3]\begin{eqnarray}}
\newcommand{\ea}{\end{eqnarray}}

\newcommand{\f}{\frac}

\def \o{\omega}

\def \O{\Omega}

\def \d{\Delta}

\def \th{\theta}

\begin{document} 

\title{Inner-most stable circular orbits in extremal and non-extremal Kerr-Taub-NUT spacetimes}
\author{Chandrachur Chakraborty}
\email{chandrachur.chakraborty@saha.ac.in}
\affiliation{Saha Institute of Nuclear Physics, Kolkata 700064, India}

\begin{abstract}
We study causal geodesics in the equatorial plane of the extremal
Kerr-Taub-NUT spacetime, focusing on the 
Innermost Stable Circular Orbit (ISCO), and compare its behaviour
 with extant results for the ISCO in the extremal Kerr spacetime. 
Calculation of  
the radii of the direct ISCO, its Kepler frequency, 
and rotational velocity show that the ISCO 
coincides with the horizon in the exactly extremal situation. 
We also study geodesics in the strong {\it non}-extremal limit, 
i.e., in the limit of vanishing Kerr parameter (i.e., for Taub-NUT
 and massless Taub-NUT spacetimes as special cases 
of this spacetime). It is shown that the radius of 
the direct ISCO increases with NUT charge 
in Taub-NUT spacetime. As a corollary, it is shown that there is 
no stable circular orbit in massless
 NUT spacetimes for timelike geodesics. 
\end{abstract}

\maketitle

\section{Introduction}

 It is perhaps Lynden-Bell and Nouri-Zonoz \cite{lnbl} 
who are the first to motivate investigation on the observational 
possibilities  for (gravito)magnetic monopoles. It has been 
claimed that signatures of such spacetimes might be found in 
the spectra of supernovae, quasars, or active galactic nuclei.
 The authors of \cite{kag} have recently brought this into focus, 
by a careful and detailed analysis of geodesics in such spacetimes.
 Note that (gravito)magnetic monopole spacetimes with angular 
momentum admit relativistic thin accretion disks of a black 
hole in a galaxy or quasars\cite{liu}. This provides a 
strong motivation for studying geodesics in such spacetimes 
because they will affect accretion in such spacetimes from massive
stars, and might offer novel observational prospects.

We know that the marginally stable orbit (also 
called Inner-most stable circular orbit (ISCO))
plays an important role in the accretion disk theory. 
That fact is important for spectral analysis of X-ray sources
 \cite{jh,abr}. The circular orbits with $r > r_{ISCO}$ 
turn out to be stable, while those with $r < r_{ISCO}$ are not.
Basically, accretion flows of almost free matter (stresses 
are insignificant in comparison with gravity
or centrifugal effects), resemble almost circular motion for 
$r > r_{ISCO}$, and almost radial free-fall for
$r < r_{ISCO}$. In case of thin disks, this transition in 
the character of the flow is expected to produce an
effective inner truncation radius in the disk. The exceptional 
stability of the inner radius of the X-ray binary LMC X-3 \cite{lmc},
 provides considerable  evidence for such a connection and, hence, 
for the existence of the ISCO. The 
transition of the flow at the ISCO may also show up in the observed 
variability pattern, if variability
 is modulated by the orbital motion \cite{abr}. One  
may expect that the there will be no 
variability observed with frequencies $\O > \O_{ISCO}$, 
i.e., higher than the Keplerian orbital frequency 
at ISCO, or that the quality factor for variability, 
$Q\sim\f{\O}{\Delta \O}$ will significantly drop at $\O_{ISCO}$.
Several variants of this idea have been discussed 
in the following references\cite{bar1,bar2}.

 The Taub-NUT geometry \cite{taub,nut} 
possesses gravitomagnetic monopoles. Basically, 
this spacetime is
a stationary and spherically symmetric vacuum solution
 of Einstein equation. As already mentioned, the authors of ref.\cite{kag} 
have made a complete classification of geodesics 
in Taub-NUT spacetimes and describe elaborately the `full'
set of orbits for massive test particles. However, 
there is no specific discussion on the various 
innermost stable orbits in such 
spacetimes for null as well as timelike geodesics. 
This is the gap in the literature which we wish 
to fulfill in this paper. Our focus here is the 
the three parameter Taub-NUT version 
of the Kerr spacetime which has angular momentum, mass and 
the NUT parameter ($n$,the gravitomagnetic monopole
strength), and is a stationary, axisymmetric 
vacuum solution of the Einstein equation.
The geodesics and the orbits of the charged particles 
in Kerr-Taub-NUT (KTN) spacetimes have also been discussed by 
Miller\cite{ml}. Abdujabbarov\cite{abu} et. al. 
discuss some aspects of these geodesics in KTN spacetime,
although the black hole solution remains a bit in doubt. 
Liu et. al.\cite{liu} have also obtained the geodesic equations but 
there are no discussions about the ISCOs in KTN spacetime. 
We know that ISCO plays many important roles in astrophysics as
well as in gravitational physics, hence the strong physical motivation to study them. 

The presence of NUT parameter lends the 
Taub-NUT spacetime a peculiar character and makes the NUT 
charge into a quasi-topological parameter.
For example in the case of maximally rotating Kerr spacetime (extremal Kerr) 
where we can see that the timelike 
circular geodesics and null circular geodesics coalesce into
a zero energy trajectory. This result and also the geodesics of the 
extremal Kerr spacetime have been elaborately described in ref. \cite{pm}.
 They show that the ISCOs
of the extremal Kerr spacetime for null geodesics and timelike 
geodesics coincide on the horizon (at $r=M$) which means that
the geodesic on the horizon must coincide with the principal null geodesic generator.
This is a very peculiar feature of extremal  spacetimes.

In the case of the non-extremal Kerr spacetime Chandrasekhar
\cite{ch} presents complete and detailed discussion on timelike 
and null geodesics (including ISCOs and other circular orbits). However, 
in that reference, subtleties associated with the extremal limit and 
special features of the precisely extremal Kerr spacetime have 
not been probed significantly. Likewise, the behaviour of geodesics,
 especially those close to the horizon (in the equatorial plane, 
where subtleties regarding geodesic incompleteness \cite{kag} 
can be avoided), have not been considered in any detail.

We wish to investigate in this paper the differences of ISCOs in Kerr-Taub-NUT
spacetimes due to the inclusion of the NUT parameter in Kerr spacetime.
While centering on the set of ISCOs, we also probe the Keplerian 
orbital frequencies and other important astronomical observables 
(namely, angular momentum ($L$), energy ($E$), rotational velocity
($v^{(\phi)}$) etc.), relevant for accretion disk physics,
 in the Kerr-Taub-NUT and Taub-NUT spacetimes; these have not been investigated 
extensively in the extant literature. The presence of the NUT parameter 
in the metric always throws up some interesting phenomena in 
these particulars spacetimes (Kerr-Taub-NUT and Taub-NUT), 
arising primarily from their topological properties, which we wish 
to bring out here in the context of our investigation on the ISCOs.
 E.g., while it has been demonstrated in ref. \cite{kag} that the 
orbital precession of spinless test particles in Taub-NUT 
geodesics vanishes, investigation by me and P. Majumdar 
has shown that in both the massive and massless Taub-NUT spacetimes 
spinning gyroscopes exhibit nontrivial frame-dragging (Lense-Thirring)
 precession \cite{cm}. Even though we do not discuss inertial 
frame-dragging in detail in this paper, this does provide 
a theoretical motivation as well to probe ISCOs in Kerr-Taub-NUT 
spacetimes, following well-established techniques \cite{ch}.

A word about our rationale for explicitly restricting to
 causal geodesics on the {\it equatorial} plane : these 
geodesics are able to avoid issues involving geodesic 
incompleteness exhibited in the spacetimes for other polar planes.
 Since there is a good deal of discussion of these issues
 elsewhere (see \cite{kag} for a competent review), we 
prefer to avoid them and concentrate instead on other issues of interest.

The paper is organized as follows : in section II we
discuss the Kerr-Taub-NUT metric and its geodesics on 
the equatorial plane of precisely extremal spacetime.
 We discuss ISCOs and other astronomical observables for 
the null geodesics and timelike geodesics in the
 section III and IV, respectively.
In each section of sections III and IV, there 
are two subsections in which we derive the radii of 
ISCOs and some other important astronomical
 observables (like $L, E, v^{(\phi)}, \O$) for Taub-NUT and 
massless Taub-NUT spacetimes as the special cases of
Kerr-Taub-NUT spacetimes. We end in section V with a
 summary and an outlook for future work.

\section{Kerr-Taub-NUT metric}  

The Kerr-Taub-NUT spacetime is geometrically stationary, axisymmetric
vacuum solution of Einstein equation with Kerr parameter $(a)$ and
NUT charge or dual mass $(n)$. This dual mass is an intrinsic
feature of general relativity. If NUT charge
\cite{nut, taub} vanishes, the solution reduces to
Kerr geometry\cite{rk}. The Kerr-Taub-NUT metric, represented here 
in Schwarzschild-like coordinates\cite{ml},
\begin{equation}
ds^2=-\f{\d}{p^2}(dt-A d\phi)^2+\f{p^2}{\d}dr^2+p^2 d\th^2
+\f{1}{p^2}\sin^2\th(adt-Bd\phi)^2
\label{lnelmnt}
\end{equation}
with 
\begin{eqnarray}\nonumber
\d&=&r^2-2Mr+a^2-n^2,  p^2=r^2+(n+a\cos\th)^2,
\\
A&=&a \sin^2\th-2n\cos\th, B=r^2+a^2+n^2.
\end{eqnarray}
At first, we want to study the geodesic motions in the equatorial
plane of the Kerr-Taub-NUT (KTN) spacetimes. So, 
the line element of the Kerr-Taub-NUT spacetime at the equator will be
\begin{equation}
 ds_e^2=-\f{r^2-2Mr-n^2}{p_e^2}dt^2-\f{4a(Mr+n^2)}{p_e^2}d\phi dt
+\f{B^2-a^2\d}{p_e^2}d\phi^2+\f{p_e^2}{\d}dr^2
\end{equation}
where,
\begin{eqnarray} 
p_e^2=r^2+n^2
\end{eqnarray}
From the above metric we can easily derive the velocity components 
(for $\th=\f{\pi}{2}$ and $\dot{\th}=0$) of the massive test particle
\cite{cp}
\begin{eqnarray}
\dot{t}=\f{1}{p^2}\left[\f{B}{\d}P(r)+aO(\th)\right]
\label{tdot}
\\
\dot{\phi}=\f{1}{p^2}\left[\f{a}{\d}P(r)+O(\th)\right]
\label{pdot}
\\
\dot{r}^2=\f{\d}{p^2}\left[k+\f{1}{p^2}\left(\f{P^2}{\d}-O^2\right)\right]
\label{rdot}
\end{eqnarray}
where,
\begin{eqnarray}
 P(r)=BE-La
\\
O(\th)=L-aE=x
\end{eqnarray}
We may set without loss of generality, 
\begin{eqnarray}\nonumber
k&=&-1 \,\, \text{for timelike geodesics} 
\\
&=&0 \,\, \text{for nulllike geodesics}
\end{eqnarray}
$L$ is specific angular momentum or angular momentum per unit mass of the test particle
and, $E$ is to be interpreted as the specific energy or 
energy per unit mass of the test particle for the timelike geodesics $k=-1$.

\section{ISCOs for the null geodesics in Kerr-Taub-NUT spacetimes}
In this section, we derive the radii of ISCOs and other important 
astronomical quantities for null geodesics $(k=0)$ in the two
spacetimes, one is Kerr-Taub-NUT spacetimes and another is
Taub-NUT spacetimes.

As we have noted, $k=0$ for null geodesics and the radial eqn. 
(\ref{rdot}) becomes 
\begin{equation}
\dot{r}^2=E^2+2(L-aE)^2\f{(Mr+n^2)}{(r^2+n^2)^2}-\f{(L^2-a^2E^2)}{(r^2+n^2)}
\label{radnull}
\end{equation}
It will be more convenient to distinguish the geodesics by the impact
parameter
\begin{equation}
 D=\f{L}{E}
\end{equation}
rather than $L$.
\\
We first observe the geodesics with the impact parameter
\begin{equation}
 D=a \,\,\, \text{or}\,\,\, L=aE
\end{equation}
Thus, in this case, equations (\ref{tdot}), (\ref{pdot}) and (\ref{radnull})
reduce to
\begin{eqnarray}
\dot{r}&=&\pm E
\\
\dot{t}&=&\f{(r^2+n^2+a^2)}{\d}E 
\label{ntdot}
\\ 
\text{and},&& \nonumber
\\
\dot{\phi}&=&\f{a}{\d}E
\label{nphidot}
\end{eqnarray}
The radial co-ordinate is described uniformly with respect to the 
affine parameter while the equations governing $t$ and $\phi$ are
\begin{eqnarray}\nonumber
 \f{dt}{dr}&=&\pm \f{(r^2+n^2+a^2)}{\d}
\\
\text{and} \,\,\,\,\,\,\,\, &&\nonumber
\\
\f{d\phi}{dr}&=&\pm \f{a}{\d}
\label{rp}
\end{eqnarray}
The solutions of these equations are

\begin{eqnarray}
 \pm t &=& r+\f{r_+^2+a^2+n^2}{r_+-r_-}\ln\left(\f{r}{r_+}-1\right)
-\f{r_-^2+a^2+n^2}{r_+-r_-}\ln\left(\f{r}{r_-}-1\right)
\\
\pm \phi&=& \f{a}{r_+-r_-}\ln\left(\f{r}{r_+}-1\right)
-\f{a}{r_+-r_-}\ln\left(\f{r}{r_-}-1\right)
\end{eqnarray}
These solutions exhibit the characteristic behaviors of $t$ and $\phi$
of tending to $\pm \infty$ as the horizons at $r_+$ and $r_-$ are approached.
 The coordinate $\phi$, like the coordinate
$t$, is not a `good' coordinate for describing what really happens
with respect to a co-moving observer: a trajectory approaching 
the horizon (at $r_+$ or $r_-$) will spiral round the spacetime
an infinite number of times even as it will take an infinite 
coordinate time $t$ to cross the horizon and neither will be experience 
of the co-moving observer.

The null geodesics which are described by eqn.(\ref{rp}),
are the members of the principal null congruences and these
are confined in the equatorial plane.

In general it is clear that there is a critical value of the 
impact parameter $D=D_c$ for which the geodesic equations 
allow an unstable circular orbit of radius $r_c$. In case of $D<D_c$,
only one kind of orbits are possible: 

it will be arriving
from infinity and cross both of the horizons and terminate
into the singularity. For $D>D_c$, we can get two types of orbits:

(a) arriving from infinity, have perihelion distances greater 
than $r_c$, terminate into the singularity at $r=0$ and $\th=\f{\pi}{2}$

(b) arriving from infinity, have aphelion distances less than
$r_c$, terminate into the singularity at $r=0$ and $\th=\f{\pi}{2}$.
For $n=0$, we can recover 
the results in the case of Kerr geometry (see eqn.(77)  of \cite{ch}).
\\ 
The equations determining the radius ($r_c$) of the stable circular 
`photon orbit' are (eqn. \ref{radnull})
\begin{eqnarray}
&& E^2+2(L-aE)^2\f{(Mr_c+n^2)}{(r_c^2+n^2)^2}-\f{(L^2-a^2E^2)}{(r_c^2+n^2)}=0
\label{null1}
\\
\text{and}\,\,\, && \nonumber
\\
&&-(L-aE)^2\f{(3Mr_c^2+4r_cn^2-Mn^2)}{(r_c^2+n^2)^3}+\f{r_c(L^2-a^2E^2)}{(r_c^2+n^2)^2}=0
\label{null2}
\end{eqnarray}
Substituting $D=\f{L}{E}$ in the above equations(\ref{null1},\ref{null2}),
we get from eqn.(\ref{null2}) 
\begin{equation}
 \f{D_c-a}{D_c+a}=\f{r_c(r_c^2+n^2)}{r_c(3Mr_c+4n^2)-Mn^2}
\label{dc1}
\end{equation}
Letting
\begin{equation}
 y=D_c+a
\end{equation}
and substituting it in eqn.(\ref{dc1}) we get
\begin{eqnarray}
 y=2a\left[1-\f{r_c(r_c^2+n^2)}{r_c(3Mr_c+4n^2)-Mn^2}\right]^{-1}
\end{eqnarray}
So, the critical value of the impact parameter $D_c$ in KTN spacetime is
\begin{eqnarray}
 D_c=2a\left[1-\f{r_c(r_c^2+n^2)}{r_c(3Mr_c+4n^2)-Mn^2}\right]^{-1}-a
\end{eqnarray}
To get the value of $r_c$ we have to take eqn.(\ref{null1})
which reduces to 
\begin{eqnarray}
 1+2(D_c-a)^2\f{(Mr_c+n^2)}{(r_c^2+n^2)^2}-\f{(D_c^2-a^2)}{(r_c^2+n^2)}=0
\label{dc2}
\end{eqnarray}
Substituting the value of
\begin{eqnarray}
 D_c=y-a=2a\left[1+\f{r_c(r_c^2+n^2)}{r_c(3Mr_c+4n^2)-Mn^2}\right]-a
\end{eqnarray}
(neglecting the higher order terms of $r_c$, we mean $r_c<n$ and 
$r_c<M$) in the eqn.(\ref{dc2})
we get the equation to determine the radius ($r_c$) of the stable circular 
photon orbit:
\begin{eqnarray}
 (9-4a_M^2)r_{cM}^4+4(6n_M^2-a_M^2)r_{cM}^3+2n_M^2(8n_M^2-6a_M^2-3)r_{cM}^2
+4n_M^2(a_M^2-2n_M^2)r_{cM}+n_M^4=0
\label{i1}
\end{eqnarray}
where the lower index `M' of a particular parameter represents
that the parameter is divided by `M' (such as, $a_M=\f{a}{M}$)
For $r_c<n$ we get the solution of $r_c$ as
\begin{eqnarray}
r_{cM}=\f{n_M}{6(a_M^2-6n_M^2)}\left[(U-18Vn_M^2+3n_M^2V)^{\f{1}{3}}
+W(U-18Vn_M^2+3n_M^2V)^{-\f{1}{3}}
+n_M(8n_M^2-6a_M^2-3)\right]
\label{isr1}
\end{eqnarray}
where the values of $U, \, V, \, W$ are given in the footnote
\footnote{\begin{eqnarray}\nonumber
U&=&-1584n_M^5a_M^2+684n_M^3a_M^4-108n_Ma_M^6-54n_M^3a_M^2-27a_M^4n_M+1152n_M^7
\\ \nonumber
&+&540n_M^5+512n_M^9-1152n_M^7a_M^2+864n_M^5a_M^4-216n_M^3a_M^6-27n_M^3
\\ \nonumber
V&=&(-192a_M^8-432a_M^8n_M^2+1224n_M^2a_M^6+1584n_M^4a_M^6-2808n_M^4a_M^4-351n_M^2a_M^4
\\ \nonumber
&-&1920n_M^6a_M^4+1080n_M^4a_M^2+1824n_M^6a_M^2+768n_M^8a_M^2-108n_M^6-162n_M^4)^{\f{1}{2}}
\\ \nonumber
W&=&(-60n_M^2a_M^2+12a_M^4+96n_M^4+64n_M^6-96n_M^4a_M^2+36n_M^2a_M^4+9n_M^2)
\end{eqnarray}}. Here, the expression of $r_{cM}$ is the radius of the ISCO as it is the
smallest positive real root of the ISCO equation(\ref{i1}).
\\

{\bf Extremal case:}
We know that in KTN spacetime, the {\it horizons} are at 
\begin{equation}
 r_{\pm}=M \pm \sqrt{M^2+n^2-a^2},
\end{equation}
where $r_+$ and $r_-$ define the event horizon and
Cauchy horizon, respectively. In case of extremal KTN
spacetime,
\begin{eqnarray}\nonumber
 r_+&=&r_-
\\ \text{or,} \,\,\,\,\,\,\,
a^2&=&M^2+n^2.
\end{eqnarray}
So, the horizon of extremal KTN spacetime is at
\begin{equation}
 r=M
\label{r=M}
\end{equation}
Now, we can substitute the value of eqn.(\ref{dc1}) 
in eqn.(\ref{null1}) (also take $a^2=M^2+n^2$) and get
the equation for determining the radius $(r_c)$ of the stable 
circular photon orbit is 
\begin{equation}
 r_c^2(r_c^2-3Mr_c-3n^2)^2-4(M^2+n^2)r_c^2(Mr_c+2n^2)+2Mn^2r_c(r_c^2-3Mr_c-n^2+2M^2)+M^2n^4=0
\label{phorbt}
\end{equation}
Interestingly, the solution of this sixth order {\it non-trivial} equation 
is
\begin{equation}
 r_c=M
\end{equation}
This is the radius of the ISCO as it is the
smallest positive real root of the ISCO equation(\ref{phorbt}).
It means that the radius of the ISCO ($r_c$) coincides 
with the horizon in extremal KTN spacetime for null geodesics. 
This is the same thing which happens in extremal Kerr spacetime also.
In extremal Kerr spacetime, the direct ISCO coincides with the horizon
at $r=M$ for null geodesic.
Now, we can calculate the critical value of the impact 
parameter in extremal KTN spacetime:
\begin{equation}
 D_c=M\f{3M^2+5n^2}{M^2+3n^2}
\end{equation}
The physical significance of the impact parameter
$D_c$ and the ISCO  have already discussed in detail.
Now, we can give our attention to the Taub-NUT
spacetime which is quite interesting.

\subsection{ISCOs in Taub-NUT spacetimes}

The Taub-NUT spacetime is a stationary and spherically
symmetric vacuum solution of Einstein equation\cite{msnr}.
Now, we are discussing a very special case in where we set
$a=0$, the angular momentum of the spacetime vanishes. We note
that for $a=0$, the primary metric (eqn. \ref{lnelmnt}) of 
Kerr-Taub-NUT spacetime reduce to the Taub-NUT metric in which 
the constant set as $C=0$. If we take the Taub-NUT 
metric in more general form, it would be
\begin{equation}\nonumber
 ds^2=-f(r)\left[dt-2n(\cos\th+C)d\phi\right]^2+\f{1}{f(r)}dr^2
+ (r^2+n^2)(d\th^2+\sin^2\th d\phi^2)
\end{equation}
where, $C$ is an arbitrary real constant. If we take $C=0$ for
the above metric,
the form of Taub-NUT metric is same as the metric (\ref{lnelmnt})
of `Kerr-Taub-NUT spacetimes with $a=0$'. Physically, $C=0$ leads
to the only possibility for the NUT solutions to have a finite total 
angular momentum\cite{manko}. After noting the above points, we write
the line element of the Taub-NUT spacetimes at the equator as
\begin{equation}
  ds_e^2=-\f{r^2-2Mr-n^2}{r^2+n^2}dt^2
+\f{r^2+n^2}{r^2-2Mr-n^2}dr^2+(r^2+n^2)d\phi^2
\end{equation}
To determine the radius $(r_{cn})$ of the stable circular photon
orbit, we put $a=0$ in eqn.(\ref{phorbt}) and get the following
 \begin{eqnarray}\nonumber
 r_{cn}^2(r_{cn}^2-3Mr_{cn}-3n^2)^2&+&2Mn^2r_{cn}
(r_{cn}^2-3Mr_{cn}-3n^2)+M^2n^4=0
\\
\text{or},\,\,\,\,\,\,\,\,\,\,\, \nonumber
\\
r_{cn}^3&-&3Mr_{cn}^2-3n^2r_{cn}+Mn^2=0
\label{i2}
\end{eqnarray}
Solving the above equation we get,
\begin{eqnarray}
r_{cn}&=&M+2(M^2+n^2)^{1/2}\cos\left[\f{1}{3}\tan^{-1}\left(\f{n}{M}\right)\right]
\\
\text{or},\,\,\,\,\,\,\,\,\,\,\, \nonumber
\\
r_{Mcn}&=&1+2(1+n_M^2)^{1/2}\cos\left[\f{1}{3}\tan^{-1}\left(n_M\right)\right]
\end{eqnarray}
This is the radius of the ISCO as it is the
smallest positive real root of the ISCO equation(\ref{i2}).
For $n_M=\f{n}{M}=0$, we can recover $r_c=3M$ of Schwarzschild spacetime.
We know that the position of {\it event horizon} in Taub-NUT spacetime is:
\begin{eqnarray}
 r_+&=&M+\sqrt{M^2+n^2}
\\
\text{or},\,\,\,\,\,\,\,\,\,\,\, \nonumber
\\
r_{+M}&=&1+\sqrt{1+n^2}
\end{eqnarray}
It seems that the position of the ISCO ($r_{cn}$)
could be in timelike, nulllike or spacelike surfaces depending
on the value of NUT charge ($n_M$) in Taub-NUT spacetime. But, 
it never be happened. We cannot get any solution of $n_M$
for $r_{Mcn}=r_{+M}$. 
If we plot (FIG.1) $r_{Mcn}$ (green) and $r_{+M}$ (red) vs $n_M$ 
we can see that $r_{Mcn}$ and $r_{+M}$ are increasing continuously for $n_M\geq 0$.
They cannot intersect each other in any point. So, we can conclude that 
ISCO could not be on the horizon for any value of $n_M$.
Thus, photon orbit is always in timelike region for any real value of $n_M$ 
in Taub-NUT spacetime.
\begin{figure}
   \begin{center}
\includegraphics[width=3in]{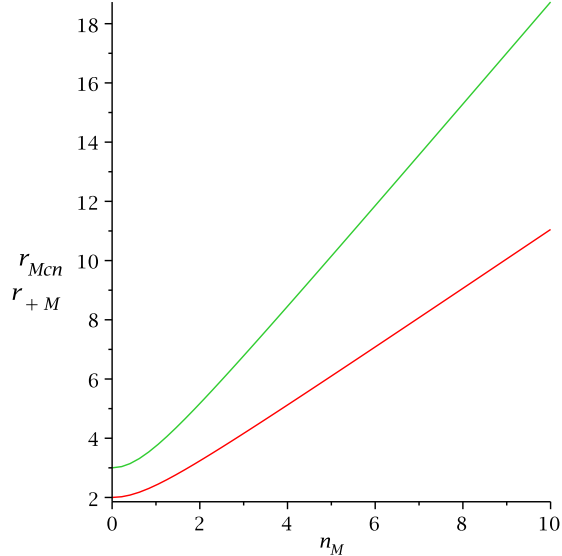}
     \caption{Plot of $r_{Mcn}$ (green) and $r_{+M}$ (red) vs $n_M$}
     \end{center}
~\label{fig_1}
\end{figure}

\subsection{ISCOs in massless Taub-NUT spacetimes}

The NUT spacetime, for the mass $M=0$ is also well-defined
(see, for example, appendix of \cite{rs}). So, we can also 
determine the radius $(r_{0cn})$ of the stable circular photon
orbit for massless Taub-NUT spacetimes. This turns out to be
\begin{equation}
 r_{0cn}=\sqrt{3}n
\end{equation}
We note that the {\it horizons} of massless Taub-NUT spacetime
are at $r=\pm n$\cite{cm}. So, the ISCO of this spacetime for null 
geodesic is always outside the {\it horizon}, we mean in the timelike surface.
Here, it should also be highlighted that in the case of Taub-NUT 
spacetime the {\it horizons} are at 
\begin{equation}
 r_{\pm}=M \pm \sqrt{M^2+n^2}
\end{equation}
as $a=0$. We can see that one {\it horizon} is located
at $r_+>2M$ and the other at $-|n|<r_-<0$.

 Thus there are mainly two important features 
of the Taub-NUT and massless Taub-NUT spacetime, one is that the 
second {\it horizon} is at a {\it negative} 
 value of the radial Schwarzschild coordinate and
another is that the both of the spacetimes 
possess the quasi-regular singularity\cite{els}.
But, the presence of the Kerr parameter $(a)$  leads to the
 {\it naked} singularity in non-extremal KTN spacetime 
as well as non-extremal Kerr spacetime. In these cases, 
both the horizons are located at 
imaginary distances. This means that we can observe the
singularity from the infinity.
So, this is called as a naked singularity. But, in the Taub-NUT spacetime
the absence of Kerr parameter never leads to any of the
horizons at imaginary distance. So, there could not be
any naked singularity in Taub-NUT and massless Taub-NUT spacetime. 
The absence of the Kerr parameter also
leads to the non-existence of any {\it extremal} case for Taub-NUT spacetime
as the {\it horizons} of Taub-NUT spacetime cannot act as
the {\it actual} horizons like the horizons of our other well-known spacetimes
(such as Kerr, Reissner-Nordstr\"om geometry). In the extremal 
KTN spacetime, the horizon is located at
$r=M$ which is same as the Kerr spacetime. We note that the NUT charge has no
any effect to determine the ISCO in extremal KTN spacetime as
the Kerr parameter takes it highest value $a=\sqrt{M^2+n^2}$
but the ISCO radius (eqn.\ref{isr1}) in non-extremal KTN spacetime depends 
on the NUT charge.

\section{ISCOs for the timelike geodesics in Kerr-Taub-NUT spacetimes}
For timelike geodesics, equations (\ref{tdot}) and (\ref{pdot})
for $\dot{\phi}$ and $\dot{t}$ remain unchanged; but equation (\ref{rdot}) 
is reduced to
\begin{equation}
 (r^2+n^2)\dot{r}^2=2(L-aE)^2\f{Mr+n^2}{r^2+n^2}+E^2(r^2+n^2)-(L^2-a^2E^2)-\d
\label{nullrdot}
\end{equation}

{\bf The circular and associated orbits:}
We now turn to consideration of the radial eqn.(\ref{nullrdot})
in general.
With the reciprocal radius $u(=\f{1}{r})$ as the independent 
variable and substituting $L-aE=x$, the equation takes the form 
\begin{eqnarray}\nonumber
&(1+n^2u^2)^2&\dot{u}^2 u^{-4}
\\
=2x^2u^3(M+n^2u)+&(1+n^2u^2)&
[E^2(1+n^2u^2)-(1+a^2u^2-n^2u^2-2Mu)-u^2(x^2+2aEx)]
\label{nulludot}
\end{eqnarray}
Like the Schwarzschild, Reissner-Nordstr\"om and Kerr geometries,
the circular orbits play an important role in the classification
of the orbits. Besides, they are useful for some 
special features of the spacetimes, after all; the reason for studying
the geodesics. 
When the values of $L$ and $E$ are completely arbitrary, the quartic
polynomial of right hand side of eqn.(\ref{nulludot}) will have a triple 
root. The conditions for the occurrence of the triple root
from the above eqn.(\ref{nulludot}) we get,
\begin{eqnarray}
2x^2u^3(M+n^2u)+(1+n^2u^2)
[E^2(1+n^2u^2)-(1+a^2u^2-n^2u^2-2Mu)-u^2(x^2+2aEx)]=0
\label{m1}
\end{eqnarray}
and
\begin{eqnarray}
3Mx^2u^2+M(1+3n^2u^2)+u(2E^2n^2-a^2)-u(x^2+2aEx)
+2n^2u^3(x^2-2aEx+E^2n^2+n^2-a^2)=0
\label{m2}
\end{eqnarray}

Equations (\ref{m1}) and (\ref{m2}) can be combined to give
\begin{eqnarray}
 E^2=1-Mu+\f{Mx^2u^3(1-n^2u^2)+2x^2n^2u^4}{(1+n^2u^2)^2}
\label{E2}
\end{eqnarray}
and
\begin{eqnarray}
2aExu(1+n^2u^2)&=&[x^2u^2(3M+3n^2u-Mn^2u^2)-ux^2]-(1+n^2u^2)[(a^2-2n^2)u-M(1-n^2u^2)]
\label{2aEx}
\end{eqnarray}
By eliminating $E$ between these equations we get the following quadratic eqn.
for $x$
\begin{eqnarray}\nonumber
&& x^4u^2\left[\left(u(3M+3n^2u-Mn^2u^2)-1\right)^2
-4a^2u^3\left(M(1-n^2u^2)+2n^2u\right)\right]\nonumber
\\
&-&2x^2u(1+n^2u^2)\left[\left(u(3M+3n^2u-Mn^2u^2)
-1\right)\left((a^2-2n^2)u-M(1-n^2u^2)\right)\nonumber
-2a^2u(1+n^2u^2)(Mu-1)\right]
\\&+&(1+n^2u^2)^2\left[(a^2-2n^2)u-M(1-n^2u^2)\right]^2=0
\label{qua}
\end{eqnarray}
The discriminant ``$\f{1}{4}(b^2-4ac)$'' of this equation is
\begin{eqnarray}
 4a^2u^3(1+n^2u^2)^2[M(1-n^2u^2)+2n^2u]\d_u^2
\end{eqnarray}
where,
\begin{eqnarray}
 \d_u=[1-2Mu+(a^2-n^2)u^2]
\end{eqnarray}
The solution of the eqn(\ref{qua}) is
\begin{eqnarray}
 u^2x^2=(1+n^2u^2)\f{Q_{\pm}\d_u-Q_+Q_-}{Q_+Q_-}=\f{(1+n^2u^2)}{Q_{\mp}}(\d_u-Q_{\mp})
\end{eqnarray}
where,
\begin{eqnarray}
 Q_+Q_-=\left[\left(u(3M+3n^2u-Mn^2u^2)-1\right)^2-4a^2u^3\left(M(1-n^2u^2)+2n^2u\right)\right]
\\
Q_{\pm}=1-u(3M+3n^2u-Mn^2u^2)\pm 2a\sqrt{u^3[M(1-n^2u^2)+2n^2u]}
\end{eqnarray}
We verify that,
\begin{eqnarray}
\d_u-Q_{\mp}=u\left[a\sqrt{u}\pm\sqrt{M(1-n^2u^2)+2n^2u}\right]^2
\end{eqnarray}
So, the solution of the $x$ takes the simple form,
\begin{equation}
 x=-\sqrt{\f{1+n^2u^2}{uQ_{\mp}}}\left[a\sqrt{u}\pm\sqrt{M(1-n^2u^2)+2n^2u}\right]
\end{equation}
It will appear presently that the upper sign in the foregoing eqn.
applies to the retrograde orbits, while the lower sign applies
to the direct orbits. We adhere to this 
convention in this whole section.
Substituting the value of $x$ in eqn(\ref{E2}), we get
\begin{eqnarray}
 E=\f{1}{\sqrt{(1+n^2u^2)Q_{\mp}}}\left[1-2Mu-n^2u^2\mp a\sqrt{u^3[M(1-n^2u^2)+2n^2u]}\right]
\label{E}
\end{eqnarray}
and the value of $L$ which is associated with $E$ is
\begin{eqnarray}
L=aE+x=\mp\sqrt{\f{M(1-n^2u^2)+2n^2u}{u(1+n^2u^2)Q_{\mp}}}
\left[1+(a^2+n^2)u^2\pm 2a(M+n^2u)\sqrt{\f{u^3}{M(1-n^2u^2)+2n^2u}}\right]
\label{L}
\end{eqnarray}
As the manner of derivation makes it explicit, $E$ and $L$ given by 
eqn.(\ref{E}) and (\ref{L}) are {\bf Energy} and {\bf Angular 
momentum} per unit mass, of a particle describing a circular orbit
of reciprocal radius $u$. The angular velocity $\O$ follows 
from the equation
\begin{eqnarray}
 \O=\f{\dot{\phi}}{\dot{t}}=\f{d\phi}{dt}=\f{aP+O\d}{BP+aO\d}
=\f{a(BE-La)+x\d}{B(BE-La)+ax\d}
\end{eqnarray}
Substituting the values of $x$, $L$ and $E$ in the above equation
we get the angular velocity of a chargeless massive test particle
which is moving in a particular orbit of radius $R(=\f{1}{u})$ 
in Kerr-Taub-NUT spacetime :
\begin{eqnarray}
 \O_{KTN}=\f{\sqrt{u^3[M(1-n^2u^2)+2n^2u]}}
{1+n^2u^2 \mp a\sqrt{u^3[M(1-n^2u^2)+2n^2u]}}
\end{eqnarray}
Generally, $\O_{KTN}$ is also called as the {\bf Kepler frequency} in
KTN spacetime.

The {\bf Time period} $(T)$ of a massive chargeless test particle
which is rotating in a orbit of radius $R(=\f{1}{u})$ 
can also be determined from the Kepler frequency by the
simple relation between $\O$ and $T$:
\begin{eqnarray}
 T=\f{2\pi}{\O_{KTN}}
\end{eqnarray}

 For $n=0$, we can easily get the {\bf Kepler frequency}
in Kerr spacetime:
\begin{eqnarray}
 \O_{K}=\f{\sqrt{Mu^3}}{1\mp a\sqrt{Mu^3}}=\f{M^{\f{1}{2}}}{R^{\f{3}{2}}\mp aM^{\f{1}{2}}}
\end{eqnarray}
It is already noted that the upper sign is applicable for the 
retrograde orbits and the lower sign is applicable for the direct orbits.
 
The {\bf Rotational velocity} $v^{(\phi)}$ of a chargeless massive 
test particle could be determined by 
the following equation:
\begin{eqnarray}
 v^{(\phi)}=e^{\psi-\nu}(\O-\o)
\label{v}
\end{eqnarray}
where, $\psi$, $\nu$ and $\o$ is defined from the general 
axisymmetric metric\cite{ch}
\begin{eqnarray}
 ds^2=-e^{2\nu}(dt)^2+e^{2\psi}(d\phi-\o dt)^2+e^{2\mu_2}(dx^2)^2+e^{2\mu_3}(dx^3)^2
\end{eqnarray}
In the above metric $\psi$, $\nu$, $\o$, $\mu_2$ and $\mu_3$ 
are the functions of $x^2$ and $x^3$.
 In our case (for Kerr-Taub-NUT spacetimes),
\begin{eqnarray}\nonumber
 e^{2\psi}&=&\f{B^2-a^2\d}{p_e^2} 
\\
\o&=&\f{2a(Mr+n^2)}{B^2-a^2\d}\nonumber
\\
 e^{2\nu}&=&\f{1}{p_e^2}\left[(r^2-2Mr-n^2)+\f{4a^2(Mr+n^2)^2}{B^2-a^2\d}\right]
\end{eqnarray}
Now, substituting the above values in the equation(\ref{v}),
we get the {\bf Rotational velocity} of a chargeless massive test particle
which is moving in a particular orbit of radius $R(=\f{1}{u})$ 
in Kerr-Taub-NUT spacetime :
\begin{eqnarray}\nonumber
 v^{(\phi)}_{KTN}&=&\f{(B^2-a^2\d)(\O-\o)}{[(r^2-2Mr-n^2)
(B^2-a^2\d)+4a^2(Mr+n^2)^2]^{\f{1}{2}}}
\\
&=&\f{(B^2-a^2\d)(\O-\o)}{(r^2+n^2)\sqrt{\d}}\nonumber
\\
&=&\f{\mp \sqrt{u[M(1-n^2u^2)+2n^2u]}\left[1+(a^2+n^2)u^2\right]-2au^2(M+n^2u)}
{\sqrt{\d_u}\left[1+n^2u^2\mp a\sqrt{u^3[M(1-n^2u^2)+2n^2u]}\right]}
\end{eqnarray}
Here, a point, considered as describing a circular orbit 
(with the proper circumference $\pi e^{\psi}$) with an angular 
velocity $\O$ (also called Kepler frequency) in the chosen coordinate frame,
will be assigned  an angular velocity,
\begin{equation}
 e^{\psi-\nu}(\O-\o)
\end{equation}
in the local inertial frame. Accordingly, a point which is considered as 
at rest in the local inertial frame (i.e., velocity components $u^{(1)}
=u^{(2)}=u^{(3)}=0$), will be assigned an angular velocity $\o$ in the 
coordinate frame. On this account the non-vanishing of $\o$ is said to 
describe as `dragging of inertial frame'. In weak gravity regime
\begin{equation}
 \o\sim\f{2J}{r^3}
\end{equation}
where $J$ is the angular momentum of the spacetime. For exact calculation 
of $\o$ in strong gravity regime, see Chakraborty and Majumdar\cite{cm}.
\\

We note that the {\bf Effective potential} expression of a massive test particle
plays many interesting roles in Gravitational Physics
 and also in Astrophysics. Here, we need this
potential at this moment to determine the radius of the ISCO in KTN spacetime.
We know that
\begin{equation}
 \f{E^2-1}{2}=\f{1}{2}\dot{r}^2+V_{eff}(r,E,L)
\label{ep0}
\end{equation}
where the effective potential governing the radial motion is 
\begin{equation}
 V_{eff}(r,E,L)=\f{1}{2}\left[(E^2-1)-\f{P^2-(r^2+n^2+O^2)\d}{(r^2+n^2)^2}\right]
\label{ep}
\end{equation}
For $n=0$, the effective potential reduces to eqn.(15.20) of \cite{jh}.
This is applicable to Kerr geometry. An important difference is 
that the potentials are energy and angular momentum dependent in
stationary spacetimes, i.e., Kerr, Kerr-Taub-NUT etc. This is not
happened in the static spacetimes, i.e, Schwarzschild, Reissner-Nordstr\"om etc.
In the static spacetimes the Effective potential is only depends on 
radial coordinate $r$, not depends on $L$ and $E$. This difference
is arisen due to the involvement of `rotational motion' in stationary
spacetimes. For example, particles that fall from infinity rotating 
in the same direction as the spacetime (positive values of $L$) move
in a different effective potential than initially counter-rotating
particles (negative values of $L$). These differences reflect, in part,
the rotational frame dragging of the spinning spacetime and the test
particles are dragged by its rotation.

Many interesting properties of the orbits of particles in the
equatorial plane could be explored with the radial equation(\ref{ep})
and the equations of the other components (which are already discussed)
of the four velocity. We could calculate the radii of circular orbits,
the shape of bound orbits etc. These are all different, depending upon
whether the particle is rotating with the black hole (corotating)
or in the opposite direction (counterrotating). For instance, in
the geometry of an extremal Kerr black hole $(a=M)$, there is a
corotating stable circular particle orbit at $r=M$ (direct ISCO) and a 
counterrotating stable circular orbit at $r=9M$ (retrograde ISCO).
However, as we already mentioned, for an introductory discussion
it seems appropriate not to all these interesting properties but
rather to focus on the one property which is most important for
Astrophysics, mainly for Accretion mechanism----the binding energy 
of the innermost stable circular particle orbit ({\bf ISCO}).

For a particle to describe a circular orbit at radius $r=R$, its initial radial 
velocity must vanish. Imposing this condition we get from eqn.(\ref{ep0})
\begin{equation}
 \f{E^2-1}{2}=V_{eff}(R,E,L)
\label{ep2}
\end{equation}
To stay in a circular orbit the radial acceleration must also vanish.
Thus, differentiating eqn.(\ref{ep0}) with respect to $r$ leads to the 
condition:
\begin{equation}
 \f{\partial V_{eff}(r,E,L)}{\partial r}|_{r=R}=0
\label{ep3}
\end{equation}
Stable orbits are ones for which small radial displacements away from
$R$ oscillate about it rather than accelerate away from it. Just as in 
newtonian mechanics, that is the condition that the effective potential 
must be a minimum:
\begin{equation}
 \f{\partial^2 V_{eff}(r,E,L)}{\partial r^2}|_{r=R}>0
\label{ep4}
\end{equation}
Equations (\ref{ep2}-\ref{ep4}) determine the ranges of $E,\, L, \,R$
allowed for stable circular orbits in the KTN spacetime. At the 
ISCO, the one just on the verge of being unstable--(eqn.{\ref{ep4})
becomes an equality. The last three equations is solved to obtain
the values of $E,\, L, \,R=r_{ISCO}$ that characterized the orbit.
Equations(\ref{ep2}) and (\ref{ep3}) have already been solved for KTN spacetime
at the very beginning of this section as the equations(\ref{m1})
and (\ref{m2}) to obtain the exact expressions of $E$ and $L$.
Now, we obtain the following from eqn.(\ref{ep4})
\begin{eqnarray}\nonumber
(r^2&+&n^2)[(a^2+2aEx)(3r^2-n^2)-2Mr^3+6Mrn^2-6n^2r^2+2n^4]
\\
&+&3x^2(r^4-4Mr^3-6r^2n^2+4Mrn^2+n^4)=0
\end{eqnarray}
Substituting the values of $Ex$ and $x^2$ in terms of $r=R$  in the above equation
we get,
\begin{eqnarray}\nonumber
&&2M^2R(2n^2R^2-3n^4-3R^4)+M\left[(R^6-n^6)-15R^2n^2(R^2-n^2)+a^2(n^4+6n^2R^2-3R^4)\right]
\\
&\mp& 8a\left[MR(R^2-n^2)+2n^2R^2\right]^{\f{3}{2}}-8n^2R^3(a^2+2n^2)=0
\end{eqnarray}
This is the equation to obtain the radius of the ISCO
in non-extremal KTN spacetime for timelike geodesic. Solving the 
above equation, we can determine
the radius of {\it inner-most stable circular orbit} in the 
non-extremal Kerr-Taub-NUT spacetime. But, this equation cannot be
solved anlytically as it is actually twelfth order equation.
So, we determine the radius of ISCO in
extremal KTN spacetime. For that we can substitute
\begin{equation}
 a^2=M^2+n^2
\end{equation}
in the above equation. Defining
\begin{eqnarray}
R_M=\f{R}{M} , \,\,\, n_M=\f{n}{M} , \,\,\, ,
a_M=\f{a}{M}=\sqrt{1+n_M^2}
\end{eqnarray}
we can rewrite the extremal ISCO equation in KTN spacetime as
\begin{eqnarray}\nonumber
&&2R_M(2n_M^2R_M^2-3n_M^4-3R_M^4)+\left[(R_M^6-n_M^6)
-15R_M^2n_M^2(R_M^2-n_M^2)+(1+n_M^2)(n_M^4+6n_M^2R_M^2-3R_M^4)\right]
\\
&+& 8\sqrt{1+n_M^2}\left[R_M(R_M^2-n_M^2)+2n_M^2R_M^2\right]^{\f{3}{2}}
-8n_M^2R_M^3(1+3n_M^2)=0
\label{i3}
\end{eqnarray}
Interestingly, the solution of the above {\it non-trivial} equation is 
\begin{equation}
 R_M=1 \,\,\, \text{or,} \,\,\, R=M
\end{equation}
This is the radius of the ISCO as it is the
smallest positive real root of the ISCO equation(\ref{i3}).
So, in extremal KTN spacetime, the direct ISCO radius does not depend on
the value of NUT charge. ISCO radius is completely determined
by the ADM mass of the spacetime. We note that the {\it horizon}
in extremal KTN spacetime is at $R=M$. 
So, we can say that $R=M$, the direct ISCO is on
the {\it horizon} in extremal KTN spacetime.
It is also further observed that the timelike
circular geodesics and null circular geodesics coalesce into a
single zero energy trajectory (both are on the $R=M$)
\begin{equation}
 E=0
\end{equation}
Thus, the geodesic on the horizon must coincide with 
the principal null geodesic generator.
 The existence of a timelike circular orbit turning into the null geodesic 
generator on the 
event horizon (nulllike) is a peculiar feature of extremal KTN spacetime.
It is not only that the energy $(E=0)$ vanishes on the ISCO in KTN spacetime
but angular momentum $(L=0)$ also vanishes on that.
The same thing is also happened in case of Kerr geometry.

Substituting $n=0$, we can recover the expressions of 
$\O$, $v^{(\phi)}$ in Kerr spacetimes. These are already described 
by Chandrasekhar\cite{ch} in details. The ISCO equation in Kerr
spacetime is
\begin{eqnarray}
R^2-6MR\mp8a\sqrt{MR}-3a^2=0
\end{eqnarray}
For extremal Kerr spacetime we can put $a=M$ in the above
equation and solving the above equation we get,
\begin{eqnarray}\nonumber
 R_{directISCO}&=&M
\\
 R_{retrogradeISCO}&=&9M
\end{eqnarray}

\subsection{ISCOs in Taub-NUT spacetimes}

To get various useful expressions in Taub-NUT spacetimes,
we should first take $a=0$. We don't want to reiterate the whole 
process for Taub-NUT spacetimes. This is the
 same as in the previous 
section in which we have done things in detail
for KTN spacetimes. The {\bf Energy} per unit mass $(E_{TN})$ and
the {\bf Angular momentum} per unit mass $(L_{TN})$ 
of a chargeless massive test particle
which is moving in a particular orbit of radius $r=\f{1}{u}$ in Taub-NUT spacetime
will be
\begin{eqnarray}
 E_{TN}=\f{(1-2Mu-n^2u^2)}{\sqrt{(1+n^2u^2)Q_{TN}}}
\label{E_tn}
\end{eqnarray}
and
\begin{eqnarray}
L_{TN}=\sqrt{\f{(1+n^2u^2)[M(1-n^2u^2)+2n^2u]}{uQ_{TN}}}
\label{L_tn}
\end{eqnarray}
where,
\begin{eqnarray}
Q_{TN}=1-u(3M+3n^2u-Mn^2u^2)
\end{eqnarray}
Now, we can find the angular velocity $\O_{TN}$ 
of a chargeless massive test particle
which is moving in a particular orbit of radius $r=\f{1}{u}$ at Taub-NUT spacetime: 
\begin{eqnarray}
 \O_{TN}=\f{\sqrt{u^3[M(1-n^2u^2)+2n^2u]}}{(1+n^2u^2)}
\label{o_tn}
\end{eqnarray}
For $n=0$, we can recover the well-known {\it Kepler frequency} for 
non-rotating star whose geometry is described by Schwarzschild metric.
The {\bf Kepler frequency} which is a very useful parameter 
in Relativistic Astrophysics, is defined as (substituting $n=0$ in eqn.(\ref{o_tn}))
\begin{eqnarray}
 \O_{Kep}^2=Mu^3=\f{M}{R^3}
\end{eqnarray}
where, $M$ is the mass of the star, $R$ is the distance of the
satellite from the centre of the star and $\O_{Kep}$ is the
uniform angular velocity of the satellite, moving in a circular
orbit of radius $R$ around the star.
 
The {\bf Rotational velocity} $ v^{(\phi)}_{TN}$ 
of a chargeless massive test particle
which is moving in a particular orbit of radius $r=\f{1}{u}$ at Taub-NUT spacetime: 
\begin{eqnarray}
 v^{(\phi)}_{TN}=\sqrt{\f{u[M(1-n^2u^2)+2n^2u]}{\d_u}}
\label{v_tn}
\end{eqnarray}
Finally, the direct ISCO equation in Taub-NUT spacetimes can be expressed as
\begin{eqnarray}
2r_M(2n_M^2r_M^2-3n_M^4-3r_M^4)+\left[(r_M^6-n_M^6)-15r_M^2n_M^2(r_M^2-n_M^2)\right]
-16n_M^4r_M^3=0
\label{tnisco}
\end{eqnarray}

It is a sixth order equation which is very difficult to solve analytically. So, 
we plot the values of $n_M$ vs $r_M$. 
\begin{figure}
\begin{center}
\subfigure[ $n_M$ varies from $0$ to $15$]
{\includegraphics[width=2in]{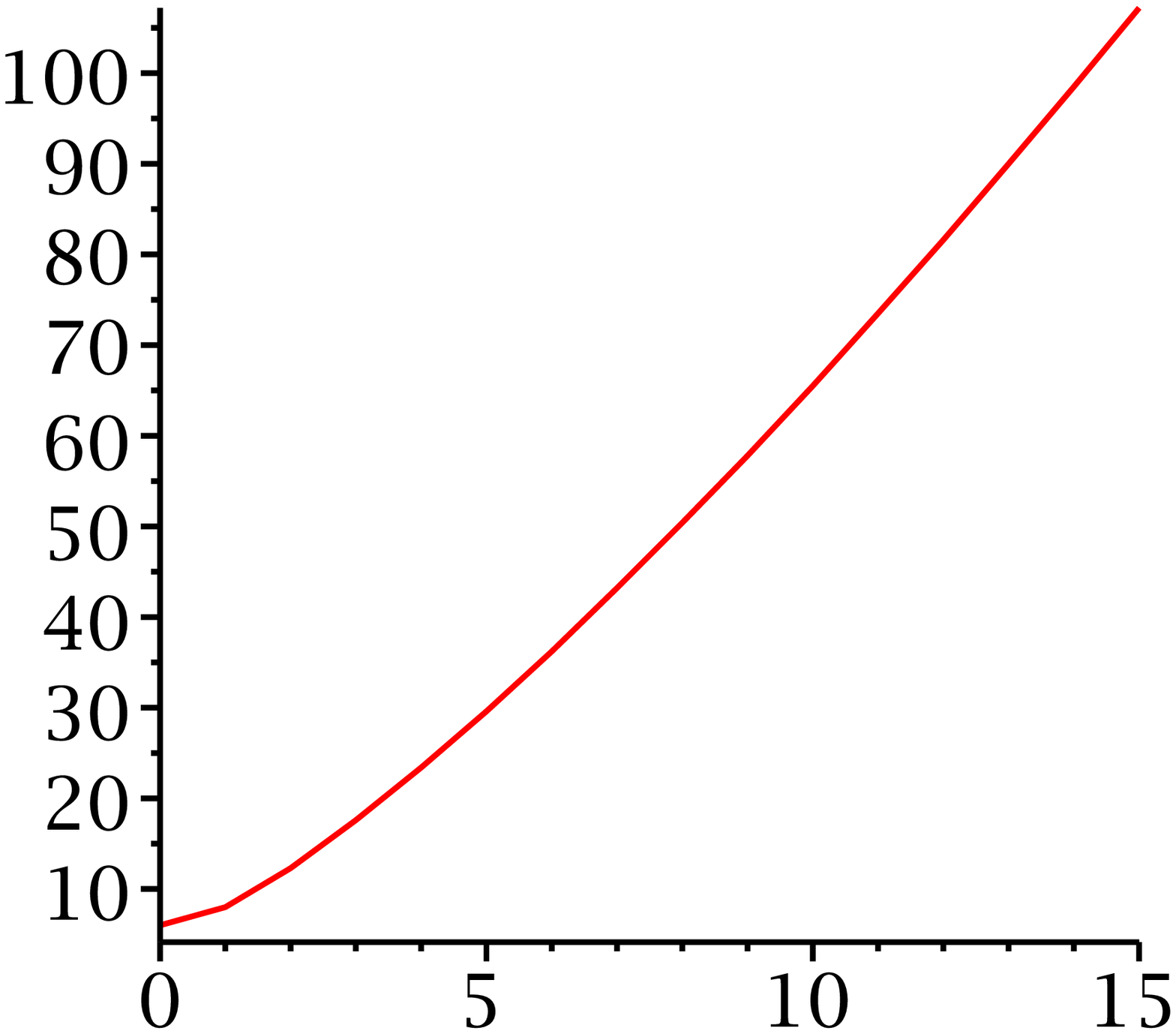}}
\subfigure[$n_M$ varies from $0$ to $100$]
{\includegraphics[width=2in]{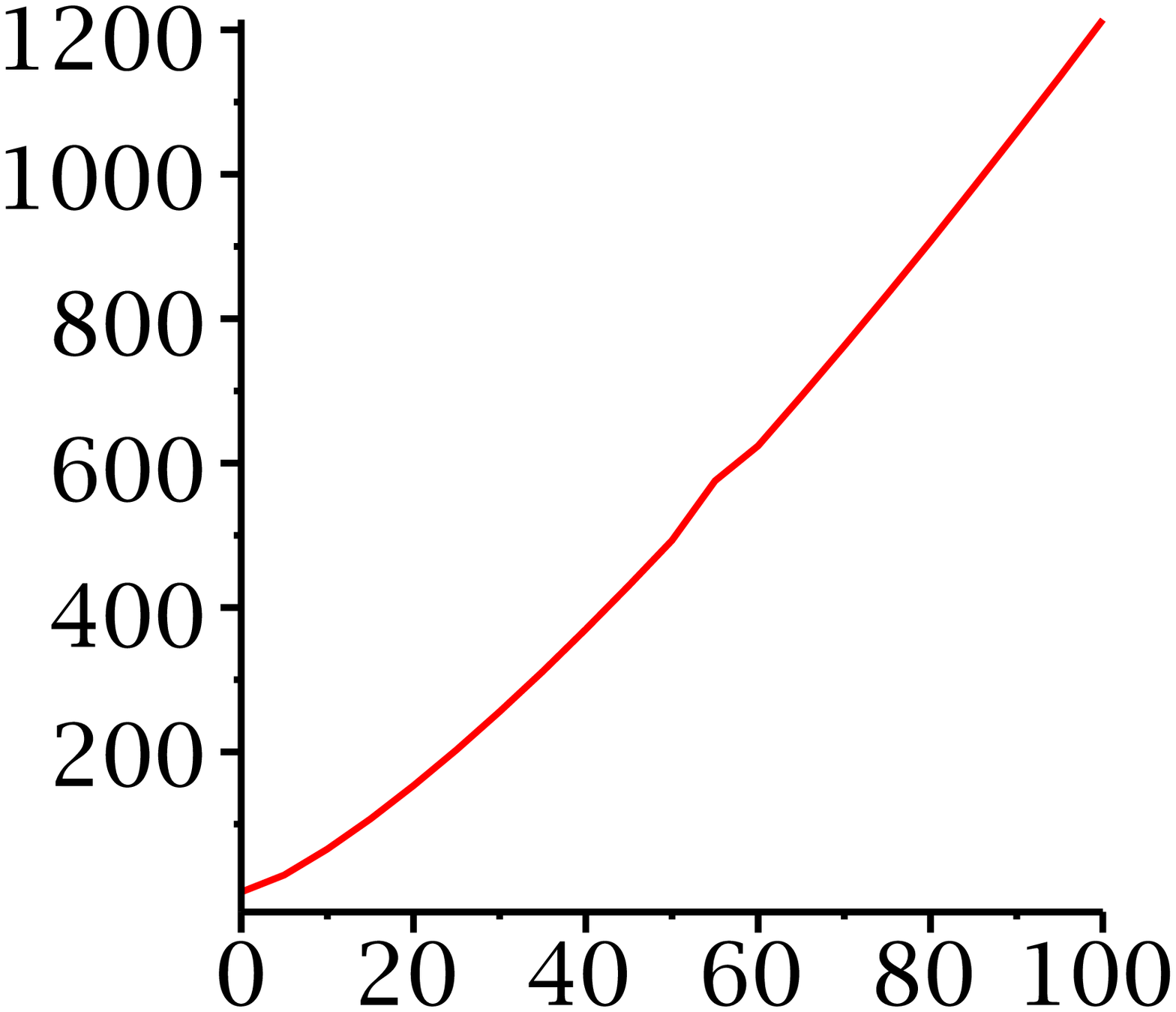}}
\caption{Plot of radius of direct ISCO $r_{Misco}$ (along $y$ axis)
vs NUT charge $n_M$ (along $x$ axis)}
\label{fig}
\end{center}
\end{figure}
 We can see from the plots (FIG. 2) and also
from the TABLE \ref{table_1} that 
in Taub-NUT spacetime, the radius of direct ISCO is increasing with 
the increasing of the NUT charge. We do not see this special 
feature in case of extremal KTN spacetimes.

\begin{table}
 \begin{tabular}{| l | c | r |  | l | c | r |}
  \hline                        
 NUT  & Radius of & Radius of 
& NUT & Radius of & Radius of 
\\
charge ($n_M$) & horizon ($r_{Mhor}$) & ISCO ($r_{Misco}$) 
& charge ($n_M$) & horizon ($r_{Mhor}$) & ISCO ($r_{Misco}$) 
 \\ 
\hline
  0(Schwarzschild) & 2 & 6  & 55 & 56 &  558 \\
\hline
  5 & 6 &  29  & 60 & 61 &  624 \\
\hline  
  10 & 11 & 65 & 65 & 66 & 693 \\
\hline
  15 & 16 & 107 & 70 & 71 &  763 \\
\hline
  20 & 21 & 153 & 75 & 76 &  834 \\
\hline
  25 & 26 & 203 & 80 & 81 &  907 \\
\hline
  30 & 31 & 256 & 85 & 86 & 982 \\ 
\hline
  35 & 36 & 312 & 90 & 91 &  1058 \\
\hline
  40 & 41 & 370 & 95 & 96 &  1135 \\
\hline
 45 & 46 & 431 & 100 & 101 &  1214 \\
\hline
  50 & 51 & 493 &  500 & 501 &  10127 \\
\hline
\end{tabular}
\caption{Comparison between radius of the horizon ($r_{Mhor}$)
and radius of the ISCO $(r_{Misco})$ at Taub-NUT spacetimes}
  \label{table_1}
\end{table}
Even though we can also solve the ISCO equation in Taub-NUT spacetimes 
with the assumption $r_M>n_M$. In this case eqn.(\ref{tnisco})
reduces to
\begin{equation}
 r_M^3-6r_M^2-15r_Mn_M^2-16n_M^4=0
\end{equation}
Solving the above equation, we get 
\begin{equation}
 r_M|_{r>n}=2+\f{1}{\alpha^{\f{1}{3}}}[\alpha^{\f{2}{3}}+5n_M^2+4]
\end{equation}
where,
\begin{equation}
 \alpha=8n_M^4+15n_M^2+8+n_M^2\sqrt{64n_M^4+115n_M^2+53}
\end{equation}
The value of $r_M$ determines the radius of direct ISCO in
Taub-NUT spacetime for timelike geodesic in case of 
the radius of the Taub-NUT spacetime is much greater
than the value of the NUT charge $n_M<r_M$.

For, $n_M=0$, we can recover $r_M=6$ as the radius of ISCO
in Schwarzschild spacetime.
\\
We can make an another assumption taking $r<n$ which helps to
reduce the eqn.(\ref{tnisco}) as
\begin{equation}
 16r_M^3-15r_M^2+6r_M+n_M^2=0
\end{equation}
Solving this eqn. we get,
\begin{equation}
 r_M|_{r<n}=\f{5}{16}+\f{1}{16\beta}[\beta^{\f{2}{3}}-7]
\end{equation}
where,
\begin{equation}
 \beta=16\sqrt{64n_M^4+115n_M^2+53}-128n_M^2-115
\end{equation}
The value of $r_M$ determines the radius of direct ISCO in
Taub-NUT spacetime for timelike geodesic in case of 
the radius of the Taub-NUT spacetime is much less
than the value of the NUT charge $n_M>r_M$.

\subsection{ISCOs in massless NUT spacetimes}

In massless NUT spacetimes, the {\bf Energy} of a chargeless massive test particle
which is moving in a particular orbit of radius $r=\f{1}{u}$ is
\begin{eqnarray}
 E_{0TN}=\f{(1-n^2u^2)}{\sqrt{(1+n^2u^2)(1-3n^2u^2)}}
\label{E_tn}
\end{eqnarray}
and, the {\bf Angular momentum} of this particle will be
\begin{eqnarray}
L_{0TN}=\sqrt{\f{2n^2(1+n^2u^2)}{(1-3n^2u^2)}}
\label{L_tn}
\end{eqnarray}
due to
\begin{eqnarray}
Q_{0TN}=(1-3n^2u^2)
\end{eqnarray}
Now, we can find the {\bf Kepler frequency} $\O_{0TN}$
of a chargeless massive test particle
which is moving in a particular orbit of radius $r=\f{1}{u}$ at massless NUT spacetime:
\begin{eqnarray}
 \O_{0TN}=\f{\sqrt{2}nu^2}{(1+n^2u^2)}
\label{O_tn}
\end{eqnarray}
and, the {\bf Rotational velocity} $v^{(\phi)}_{0TN}$  of 
this test particle at massless NUT spacetime 
will be
\begin{eqnarray}
 v^{(\phi)}_{0TN}=nu\sqrt{\f{2}{(1-n^2u^2)}}
\label{v_tn}
\end{eqnarray}
The expressions of all astronomical observables are
valid for $\f{1}{u}=r>\sqrt{3}n$ as $L$ and $E$ would be complex for 
for $r<\sqrt{3}n$ and diverges for $r=\sqrt{3}n$.
Finally, the direct ISCO equation in massless NUT spacetimes can be expressed as
\begin{eqnarray}
16n^4r^3=0
\end{eqnarray}
As $n\neq0$, we get 
\begin{eqnarray}
 r=0
\end{eqnarray}

 This means that the innermost stable circular orbit is in the 
centre of the massless NUT spacetime. This is unphysical, 
because, in the classification of of Ellis and Schmidt \cite{els, smd}
the singularity of Taub-NUT spacetime has been termed as quasiregular singularity,
since the curvature remains finite (see also \cite{gr, isr}).
As $r=n$ has the singularity in the massless Taub-NUT spacetime, all 
geodesics are terminated at $r=n$ \cite{kag} before reaching $r=0$, 
which may be regarded as a {\it spacelike} surface. So, we cannot go
beyond $r<n$ to determine ISCO. Apparently, the curvature of the
Taub-NUT spacetime shows that $r=n$ is not a singular
surface but in reality, $\th=0, \pi$ are the singularities 
in the Taub-NUT spacetimes, as the curvature
diverges for $\th=0$ and $\pi$ \cite{mkg}. Thus, the final conclusion is that
there is no any physically reliable innermost stable circular orbit in massless
NUT spacetime for timelike geodesic.

\section{Summary and Discussion}

In this paper we have calculated the radius of the innermost-stable 
circular orbit (ISCO) exactly for extremal Kerr-Taub-NUT,
Taub-NUT, massless Taub-NUT spacetimes.
We also calculated some other important astronomical observables 
(i.e., $L, E, v^{(\phi)}, \O_{Kep}$) for Kerr-Taub-NUT
 (non-extremal and extremal both), Taub-NUT and massless Taub-NUT
spacetimes. All the ISCOs of the non-extremal KTN, Taub-NUT and massless Taub-NUT 
lie on the timelike surface but the ISCOs of the extremal KTN spacetime
(both for nulllike and timelike geodesics) lie
on the horizons,i.e, on the lightlike surface.
We cannot solve the ISCO equation of the non-extremal KTN spacetime
 analytically. However, one can see intuitively from the ISCO equation
that the ISCO must belong to the timelike region, i.e., $r_{ISCO}>r_+$
where $r_+=M+\sqrt{M^2+n^2-a^2}$. The Kerr parameter ($a_k$) of
the extremal Kerr metric can take the highest value as  $a_K=M$
but it takes $a_{KTN}=\sqrt{M^2+n^2}$ in case of extremal KTN spacetime.
 So, the highest angular momentum 
of a Kerr spacetime is $J_K=M^2$ but it will be $J_{KTN}=M\sqrt{M^2+n^2}$
in the case of extremal KTN spacetime. Thus, the ratio between these
two is 
\begin{equation}
 \f{J_K}{J_{KTN}}=\f{a_K}{a_{KTN}}=\f{M}{\sqrt{M^2+n^2}} .
\end{equation}
But the interesting thing is that the radii of the ISCOs 
in extremal KTN spacetime is the same as $R=M$. Thus,
\begin{equation}
 \f{r_{KISCO}}{r_{KTNISCO}}=1
\end{equation}
 It is quite remarkable that the NUT charge 
 which appears in the expression of maximal angular momentum ($J$) 
has no effect for determining 
the radius of the ISCO in extremal KTN spacetimes.
 So, the radius of the ISCO is
independent of NUT charge $n$. We know that the radius of
the ISCO in extremal Kerr spacetime is $M$ \cite{ch} for
both null and timelike geodesics. It is expected
that if a new parameter appears in Kerr geometry, 
the ISCO should be altered away from $M$. But, this is not happening
in extremal KTN spacetime though the Kerr metric is modified with 
a new parameter $n$ as KTN spacetime. 
This non-dependence of ISCO behaviour in the extremal case is somewhat 
mysterious. The NUT parameter in extremal KTN spacetime behaves like a {\it shield},
so that the ISCO could not go away from the horizon.
Another interesting
thing is that the geodesic on the horizon coincides
with the principal null geodesic generator for both extremal
Kerr and extremal KTN spacetimes.

In the very non-extremal spacetime, when Kerr parameter vanishes,
i.e., in case of the Taub-NUT spacetime, the ISCO is shifted 
to $M+2(M^2+n^2)^{1/2}\cos\left[\f{1}{3}\tan^{-1}
\left(\f{n}{M}\right)\right]$ for null geodesics.
For timelike geodesics, the ISCO of Taub-NUT spacetime is also shifted from
null surface (in the case of extremal KTN spacetime) to timelike surface. 
The amount of this shifting is discussed in detail in the
subsection A of sections III and IV.
Another thing is that the massless NUT spacetime does not
hold any inner stable circular orbit for timelike geodesics.
But, it holds an ISCO at $\sqrt{3}n$ (which is outside the
{\it horizon}) for null geodesics. 

There are additional avenues of further work currently being 
investigated for an understanding of the thin disks accretion mechanism
in Kerr-Taub-NUT spacetime, applicable to some special 
candidates (like, supernovae, quasars, or active galactic nuclei etc.)
which are of astrophysical importance. 
One could also investigate the exact Lense-Thirring
precession rate of a test particle which is rotating on the ISCO 
in strong gravity regime of extremal KTN spacetimes. This is very 
important in accretion mechanism as accretion disk is mainly formed
near the ISCO which lies in the strong gravity regime.
Chakraborty and Majumdar\cite{cm} derived the exact Lense-Thirring (LT)
precession rates in the KTN, Taub-NUT and massless Taub-NUT
spacetimes but their formula is valid only in timelike spacetime;
it is not valid on the horizon, i.e., null surface. Thus, it is not possible 
to derive the exact LT precession rate on the ISCO
of extremal KTN spacetime by that formula. But, it would be quite helpful
for accretion mechanism in the future.
\\

{\bf Acknowledgements :} I would like to thank Prof.Dr. Parthasarathi
Majumdar for many useful discussions on this topic which have been very helpful. 
Special thanks due to Mr. Partha Pratim Pradhan for
suggesting me this problem. I would also like to
thank S. Bhattacharjee, P. Byakti and A. Majhi for 
their useful suggestions. I am grateful to the Department of Atomic
Energy (DAE, Govt. of India) for the financial assistance.

\end{document}